\documentclass[twocolumn]{revtex4}
\newcommand{\1}{\protect{1 \hspace{-0.24em} {\bf l}}}

\begin{document}

\title{Nonequilibrium Work Relation from Schr\"{o}dinger's Unrecognized Probability Theory}
\author{T. Koide}
\email{tomoikoide@gmail.com,koide@if.ufrj.br}
\affiliation{Instituto de F\'{\i}sica, Universidade Federal do Rio de Janeiro, C.P.
68528, 21941-972, Rio de Janeiro, RJ, Brazil}

%date{\today}

\begin{abstract}
Jarzynski's nonequilibrium work relation can be understood as the 
realization of the (hidden) time-generator reciprocal symmetry 
satisfied for the conditional probability function.
To show this, we introduce the reciprocal process where 
the classical probability theory is expressed with real wave functions, 
and derive a mathematical relation using the symmetry. 
We further discuss that the descriptions by the standard Markov process from an initial equilibrium state are indistinguishable from those by the reciprocal process. 
Then the Jarzynski relation is obtained from the mathematical relation for the Markov processes described by the Fokker-Planck, Kramers and relativistic Kramers equations. 
\end{abstract}

\maketitle

\section{Introduction}

In recent years the nonequilibrium behaviors of small fluctuating systems attract a great deal of interest 
\cite{sekimoto,seifert}. 
One of the main concerns is the nonequilibrium work relations \cite{jar97,bochkov77,evens93,crook99,kawasaki73,hummer01,oono98,hatano01}.
For example, Jarzynski's work relation describes the connection between 
the change of the free energy and the distribution of work for the event ensemble \cite{jar97}. 
This is satisfied for any nonequilibrium process from an initial equilibrium state, 
and its validity has been studied from various points of view \cite{jar-book,kurchan07,evens02,harris07,chet11,garcia12}.

Because of its broad applicability, the Jarzynski relation will be associated 
with not the detailed behavior of systems but a global property such as symmetry.  
In this paper, we rederive the Jarzynski relation from the perspective of symmetry satisfied for time generators of the conditional probability function.
We call it time-generator (TG) reciprocal symmetry (Eq.\ (\ref{eqn:l-ld})).
To manifest this symmetry, 
we introduce the reciprocal process which was proposed by Schr\"{o}dinger in 1931 \cite{sch31,sch32} and developed by Bernstein \cite{ber32,jam74,eqm1,eqm2,vigier,yasue,zam13,leon13}.
There, two different real wave functions are introduced and the TG reciprocal symmetry 
is associated with the exchange of those wave functions.
Then the mathematical relation, (Eq.\ (\ref{eqn:gr}) or (\ref{eqn:gr2})),  
is derived using the (hidden) TG reciprocal symmetry. 
Afterward, we show that the descriptions by the standard Markov process from an initial equilibrium state are indistinguishable from those by the reciprocal process.
Finally, using the derived mathematical relation, we obtain the Jarzynski relation  
for the Markov processes described by the Fokker-Planck equation, the Kramers equation and the relativistic Kramers equation.

The formulation developed here 
is the generalization of the previous work developed by the present author \cite{koide17}.
As the related approaches, see Refs. \cite{kurchan07,harris07,chet11,garcia12}.

This paper is organized as follows.
In Sec.\ \ref{sec:2}, we briefly summarize the reciprocal process. 
In Sec.\ \ref{sec:3}, the TG reciprocal symmetry is introduced and the mathematical relation for the reciprocal process is derived using it. 
This relation is applied to the Fokker-Planck equation, reproducing the Jarzynski relation 
in Sec.\ \ref{sec:4}.
The role of the hidden TG reciprocal symmetry is discussed 
in Sec.\ \ref{sec:5}.
Section \ref{sec:6} is devoted to concluding remarks.

In the following, $k_B$ and $c$ denote the Boltzmann constant and the speed of light, respectively.

\section{Reciprocal process}\label{sec:2}

The idea of the reciprocal process was introduced by Schr\"{o}dinger \cite{sch31,sch32} and mathematically elaborated 
by Bernstein \cite{ber32}. 
See also Refs.\ \cite{jam74,eqm1,eqm2,vigier,yasue,zam13,leon13}.
Although it is possible to formulate this theory in the more mathematically rigorous manner done by Bernstein,
we here follow the original argument by Sch\"{o}dinger, which is intuitively understandable. 

Let us consider the probability density of a stochastic (Brownian) particle.
Suppose that the initial and final probability densities 
are already known (fixed).
Then we describe the probability density at an intermediate time $t$ 
between the fixed initial and final times, $(t_i \le t \le t_f)$.

As the simplest case, we consider that the initial and final positions of the particle are given by small domains $(x_i, x_i+dx_i)$ and $(x_f, x_f+dx_f)$, respectively. 
Then the probability density of the position of the particle $x$ at $t$ is given by 
\begin{eqnarray}
h(x_f,t_f;x,t;x_i,t_i) = \frac{f_3 (x_f,t_f;x,t;x_i,t_i)}{f_2 (x_f,t_f;x_i,t_i)},
\end{eqnarray}
where $f_n (x_n,t_n; \cdots; x_1,t_1)$ represents the joint probability function of order n.
The denominator comes from the fixed initial and final probability densities, leading to 
$\int dx\ h(x_f,t_f;x,t;x_i,t_i) =1$.
This quantity $h$ plays an important role in the reciprocal process and called dual transition probability density.
In particular, assuming the Markov property for $f_n$, 
$h$ is reexpressed as  
\begin{eqnarray}
h(x_f,t_f;x,t;x_i,t_i) = \frac{f_{1|1} (x_f,t_f|x,t)f_{1|1} (x,t|x_i,t_i)}{f_{1|1} (x_f,t_f|x_i,t_i)}, \label{eqn:h-para}
\end{eqnarray}
where $f_{1|1}$ is the conditional probability function. 
The definitions of the joint probability function and the conditional probability function are the same 
as those in the standard textbook of the probability theory. 
See, for example, chapter 1 of Ref.\ \cite{breuer-book}.

When the initial and final positions are distributed, the probability density is given by 
\begin{eqnarray}
\rho_\theta (x,t) = \int dx_i dx_f\ h(x_f,t_f;x,t;x_i,t_i)c_\theta (x_f,t_f;x_i,t_i). \label{eqn:rp-probability1}
\end{eqnarray}
Here, the boundary joint probability density, $c_\theta$, represents the correlation of the initial and final probability densities, satisfying the following boundary conditions, 
\begin{eqnarray}
\int dx_f\ c_\theta (x_f,t_f;x_i,t_i) &=& \rho_\theta (x_i,t_i), \label{eqn:boundary1}\\
\int dx_i\ c_\theta (x_f,t_f;x_i,t_i) &=& \rho_\theta (x_f,t_f), \label{eqn:boundary2}
\end{eqnarray} 
where $\rho_\theta (x_i,t_i)$ and $\rho_\theta (x_f,t_f)$ are the fixed initial and final probability densities, respectively.

There are various choices of $c_\theta$ leading to 
different reciprocal processes.
If there is no correlation in the choice of the initial and final probability densities, 
$c_\theta(x_f,t_f;x_i,t_i) = \rho_\theta (x_f,t_f)\rho_\theta (x_i,t_i)$.
In this work, however, we use the form proposed by Schr\"{o}dinger, 
\begin{eqnarray}
c_\theta (x,t;y,s) = \overline{\theta}(x,t) f_{1|1} (x,t|y,s) \underline{\theta} (y,s), \label{eqn:entangle}
\end{eqnarray}
which is found by considering the optimization of the Kullback-Leibler entropy 
for $c_\theta$ and $f_{1|1}$ \cite{sch32}.
The real functions $\underline{\theta}$ and $\overline{\theta}$ play 
the role of the wave functions because, 
substituting this into Eq.\ (\ref{eqn:rp-probability1}), 
the probability density is expressed by 
\begin{eqnarray}
\rho_\theta (x,t) = \overline{\theta}(x,t)\underline{\theta} (x,t), \label{eqn:probability}
\end{eqnarray}
with the following definitions of the time evolutions, 
\begin{eqnarray}
\underline{\theta} (x,t) &=& \int dx_i\  f_{1|1} (x,t|x_i,t_i)  \underline{\theta} (x_i,t_i), \\
\overline{\theta} (x,t) &=& \int dx_f\ \overline{\theta} (x_f,t_f) f_{1|1} (x_f,t_f|x,t).
\end{eqnarray} 
We then confirm that Eqs.\ (\ref{eqn:boundary1}) and (\ref{eqn:boundary2}) 
are satisfied.

Suppose that the conditional probability function is characterized 
by the time generator $\hat{\cal L}_t$ as
\begin{eqnarray}
f_{1|1} (x,t|y,s) = \langle x | U(t,s) | y \rangle,
\end{eqnarray}
where, for $t > s$,
\begin{eqnarray}
U(t,s)
&=& 
1 + \sum_{n=1}^\infty \int^t_s dt_1 \int^{t_1}_s dt_2 \cdots \int^{t_{n-1}}_s dt_n \hat{\cal L}_{t_1} \cdots \hat{\cal L}_{t_n} \nonumber \\
&\equiv& 
T e^{\int^t_s d\tau \hat{\cal L}_\tau}.
\end{eqnarray}
Here we introduced the time-ordered product.
To express the result in a similar fashion to quantum mechanics, we introduce the bra-ket notation and the 
eigenstate of the position operator satisfying 
\begin{eqnarray}
\int dx | x \rangle \langle x | = 1 \ \ \ \ 
\langle x | x^\prime \rangle = \delta(x-x^\prime). \label{eqn:x}
\end{eqnarray}
Note that $(| x \rangle)^\dagger = \langle x |$ where $\dagger$ denotes the usual self-adjoint operation.
Then the time evolutions of the two wave functions are represented by  
\begin{eqnarray}
\partial_t | \underline{\theta} (t) \rangle &=& \hat{\cal L}_t | \underline{\theta} (t) \rangle, \label{eqn:de1}\\ 
\partial_t \langle \overline{\theta} (t) | &=& \langle \overline{\theta} (t) | (- \hat{\cal L}_t). \label{eqn:de2}
\end{eqnarray}
From Eq.\ (\ref{eqn:probability}), 
the conservation of probability leads to 
\begin{eqnarray}
\langle \overline{\theta} (t) | \underline{\theta} (t) \rangle =1.
\end{eqnarray}
Note that $(| \underline{\theta} (t) \rangle)^\dagger = \langle \underline{\theta} (t) |$, but 
$\langle \overline{\theta} (t) | \neq \langle \underline{\theta} (t) |$ in general.
The appearance of such a two-vectors formulation (bi-orthogonal system) in statistical physics 
is investigated in Ref.\ \cite{koide17}. 
For the bi-orthogonal formulation of quantum mechanics, see Ref.\ \cite{brody14}.

The expectation value of an operator $\hat{A}$ is then represented by 
\begin{eqnarray}
\langle \overline{\theta}(t) | \hat{A} | \underline{\theta}(t) \rangle 
&=& 
\langle \overline{\theta}(t_f) | 
U(t_f,t) \hat{A} U(t,t_i)  | \underline{\theta}(t_i) \rangle.
\label{eqn:exp-rp}
\end{eqnarray}
To derive the Jarzynski relation, the joint probability density should be defined 
in the reciprocal process. 
For this, we need to know the general properties satisfied for $h$ ($t_i \le v < u < t < s \le t_f$), which  
are summarized as
\begin{eqnarray}
 h(x,s;y,t;z,u) &\ge& 0, \\
 \int dy\ h(x,s;y,t;z,u) &=& 1,
\end{eqnarray} 
and the reciprocal condition,  
\begin{eqnarray}
\lefteqn{
h(w,s;x,t;y,u)h(w,s;y,u;z,v)} && \nonumber \\
&&  = h(w,s;x,t;z,v)h(x,t;y,u;z,v).
\end{eqnarray}
The reciprocal condition corresponds to the Chapman-Kolmogorov equation in the Markov process.
The parameterization by Eq.\ (\ref{eqn:h-para}) 
satisfies these properties. 
Then the joint probability density in order n is defined by 
\begin{eqnarray}
\lefteqn{r_{\theta }(x_N,t_N;x_{N-1},t_{N-1};\cdots;x_1,t_1)} && \nonumber \\ 
&& = \int dx_i d x_f\ h(x_f,t_f;x_1,t_1;x_i,t_i) h(x_f,t_f;x_2,t_2;x_1,t_1) \nonumber \\
&& \times \cdots h(x_f,t_f;x_N,t_N;x_{N-1},t_{N-1}) c_{\theta}(x_f,t_f;x_i,t_i) \nonumber \\
&&  = \int dx_i d x_f\ \overline{\theta} (x_f,t_f)
g(x_f,t_f;\cdots;x_i,t_i) 
\underline{\theta}(x_i,t_i), \label{eqn:jpd-rp}
\end{eqnarray}
where
\begin{eqnarray}
\lefteqn{g(x_f,t_f;\cdots;x_i,t_i)} && \nonumber \\
&=& f_{1|1}(x_f,t_f;x_N,t_N) 
\cdots  f_{1|1}(x_1,t_1;x_i,t_i),
\end{eqnarray}
for $t_i \le t_1 <  \cdots < t_N \le t_f$.

In the following, the expectation values with the (joint) probability densities are discussed, 
but the quantity like the probability amplitude in quantum mechanics is not considered.

\section{Mathematical relation}\label{sec:3}

The generator $\hat{\cal L}_t$ 
is not necessarily self-adjoint, $\hat{\cal L}_t \neq \hat{\cal L}^\dagger_t$ and 
can be time-dependent. However, we consider the following symmetry, 
\begin{eqnarray}
\hat{\eta}_t \hat{\cal L}_t \hat{\eta}^{-1}_t = \hat{\cal L}^\dagger_t. \label{eqn:l-ld}
\end{eqnarray}
This is a kind of pseudo-Hermitian condition considered by Dirac \cite{dirac42,pauli43} 
and has been used in the non-Hermitian Hamiltonian dynamics of quantum mechanics \cite{brody14,mostafa02, bender07}. 
The relation to the detailed balance condition used in Ref.\ \cite{kurchan07} is discussed in Sec.\ \ref{sec:6}. 
See also the discussion in  Ref.\ \cite{chet11}. 
In this paper, we call this property time-generator (TG) reciprocal symmetry.
The connection between Eq.\ (\ref{eqn:l-ld}) and symmetry is clarified only by considering the Markov process 
as a special case of the reciprocal process as is done in this paper. 
See App.\ \ref{app:t-reciprocal} where the relation between the TG reciprocal symmetry and the invariance of the Lagrangian is shown.

In the following, we consider the special case, 
\begin{eqnarray} 
\hat{\eta}_t = e^{\hat{G}_t}
\end{eqnarray} 
with the self-adjoint operator $\hat{G}_t$.
When this condition is satisfied, we can define a new self-adjoint operator by 
$e^{\hat{G}_t/2} \hat{\cal L}_t e^{-\hat{G}_t/2} = (e^{\hat{G}_t/2} \hat{\cal L}_t e^{-\hat{G}_t/2})^\dagger$. 
Then, as is done in Ref.\ \cite{koide17}, 
the eigenstates of the two generators, 
$\hat{\cal L}_t$ and $\hat{\cal L}^\dagger_t$, are shown to  
form the bi-orthogonal system and any states are expanded with the set of the eigenstates.

The eigenstates of $\hat{\cal L}_t$ and $\hat{\cal L}^\dagger_t$ are defined by  
\begin{eqnarray}
\hat{\cal L}_t | \underline{n},t \rangle &=& - \bar{\lambda}_n (t) | \underline{n},t \rangle, \label{eqn:eigen1}\\
\hat{\cal L}^\dagger_t | \overline{n},t \rangle &=& - \bar{\lambda}_n (t) | \overline{n},t \rangle, \label{eqn:eigen2}
\end{eqnarray}
satisfying 
\begin{eqnarray}
\langle \overline{n} ,t| \underline{m} ,t \rangle &=& \delta_{nm},  \label{eqn:normal}\\
\sum_n | \underline{n},t \rangle \langle \overline{n},t | &=& 1.
\end{eqnarray}
As is shown in Ref. \cite{koide17}, 
$- \bar{\lambda}_n (t)$ is even the eigenvalue of 
$e^{\hat{G}_t/2} \hat{\cal L}_t e^{-\hat{G}_t/2}$. 
It should be noted that the eigenstates are time-dependent, but are not the solutions of 
the differential equations (\ref{eqn:de1}) and (\ref{eqn:de2}).

To derive the Jarzynski relation, it is enough to consider the case of no degeneracy in the eigenstates.
Applying the TG reciprocal symmetry to Eqs.\ (\ref{eqn:eigen1}) and (\ref{eqn:eigen2}), 
we find the following relation, 
\begin{eqnarray}
e^{\hat{G}_t} | \underline{n} ,t \rangle = {\cal N}_n (t) | \overline{n} ,t \rangle, \label{eqn:wf-relation1}
\end{eqnarray}
where 
${\cal N}_n (t) = \langle \underline{n} ,t| e^{\hat{G}_t} | \underline{n} ,t \rangle$.
This is an important property in the following derivation.

Let us consider the evolution from one of the eigenstates, $| \underline{m}, t_i \rangle$. 
The corresponding final state, $\langle \overline{\theta}_m (t_f)|$, is determined to satisfy Eqs.\ (\ref{eqn:boundary1}) and 
(\ref{eqn:boundary2}), 
leading to 
\begin{eqnarray}
&& \langle \overline{\theta}_m (t_i)| = \langle \overline{m}, t_i | , \\
&& \langle \overline{\theta}_m (t_f)| U(t_f,t_i) | \underline{m}, t_i \rangle = 1. \label{eqn:thetam}
\end{eqnarray} 
Note that $U^\dagger(t,s) \neq U^{-1} (t,s)$ in general.

Now we find the following mathematical relation defined with the joint probability density, 
\begin{eqnarray}
\lefteqn{\langle \overline{\theta}_m (t_f) | e^{-\Delta G} | \underline{m} ,t_i \rangle_{jt}} && \nonumber \\
&& \hspace*{-0.7cm} \equiv 
\lim_{N\rightarrow \infty} \int dx_i dx_f 
[e^{- \dot{G}_{t_f}(x_f) dt} \prod_{j=1}^N dx_j e^{- \dot{G}_{t_j}(x_j) dt}  ] \nonumber \\
&& \times \overline{\theta}_m (x_f,t_f) g (x_f,t_f;\cdots;x_i,t_i) \underline{m}(x_i,t_i) \nonumber \\
&& \hspace*{-0.7cm} =  
\lim_{N\rightarrow \infty} 
 \langle \overline{\theta}_m  (t_f) | e^{- \dot{\hat{G}}_{t_f} dt} 
U(t_f,t_N)
e^{- \dot{\hat{G}}_{t_N} dt} 
U(t_N,t_{N-1}) \nonumber \\
&& \times e^{- \dot{\hat{G}}_{t_{N-1}} dt} 
\cdots  
e^{- \dot{\hat{G}}_{t_1} dt} 
U(t_1,t_i)
| \underline{m} , t_i \rangle \nonumber \\
&& \hspace*{-0.7cm} =
\langle \overline{\theta}_m  (t_f) | e^{-\hat{G}_{t_f}} T e^{\int^{t_f}_{t_i}ds \hat{\cal L}^\dagger_s} | \overline{m},t_i \rangle
 \langle \underline{m} ,t_i | e^{\hat{G}_{t_i}} | \underline{m} ,t_i \rangle, 
\label{eqn:gr}
\end{eqnarray}
where $dt = (t_f-t_i)/(N+1)$, $t_L = t_i + L dt$ $(1\le L \le N)$, and 
\begin{eqnarray}
\langle x | \hat{G}_t | x' \rangle = G_t (x) \delta(x-x').
\end{eqnarray}
In this derivation, we used 
\begin{eqnarray}
\int^{t_L}_{t_{L-1}} ds \hat{\cal L}_s &=& \hat{\cal L}_{t_{L-1}}(t_L - t_{L-1}), \\ 
\dot{\hat{G}}_{t_L} dt &=& \hat{G}_{t_L} - \hat{G}_{t_{L-1}},
\end{eqnarray} 
for the large $N$ limit, and Eq.\ (\ref{eqn:wf-relation1}) in the last line.
We further assumed $\hat{G}_t\hat{G}_{t'} = \hat{G}_{t'}\hat{G}_{t}$ which is satisfied for all examples discussed below.
This is the main result of this paper and more general than the Jarzynski relation. As is discussed soon later, this relation is satisfied for not only the reciprocal process and but also the standard Markov process. In fact, 
the Jarzynski relation is obtained as a special case of this general relation.

\section{Jarzynski relation}\label{sec:4}

In the standard Markov process, the time-evolution of the probability density is described by 
\begin{eqnarray}
\rho(x,t) = \int dx_i f_{1|1} (x,t;x_i,t_i) \rho(x_i,t_i), \label{eqn:markovpro}
\end{eqnarray}
for $t > t_i$. 
Although we use the same symbol for simplicity, $f_{1|1}$ here is generally different from that in the reciprocal process. 
In fact, the generator of $f_{1|1}$ here should satisfy one more condition in addition to the positivity and the Chapman-Kolmogorov equation, which is associated with the conservation of probability,  
\footnote{This is not necessarily employed for the reciprocal process. See Refs. \cite{eqm1,eqm2}.}
\begin{eqnarray}
\int dx \langle x | \hat{\cal L}_t | x^\prime \rangle = 0. \label{eqn:cond-L-ad}
\end{eqnarray} 
With this generator, Eq.\ (\ref{eqn:markovpro}) can be expressed as the differential equation, 
$\partial_t | \rho(t) \rangle = \hat{\cal L}_t | \rho(t) \rangle$ with $\langle x | \rho (t) \rangle = \rho (x,t)$.

Introducing the state $| \1 \rangle $ defined by    
\begin{eqnarray}
\langle \1 | x \rangle = \langle x | \1 \rangle = 1,
\end{eqnarray}
the expectation value in the Markov process is expressed by
\begin{eqnarray}
\int dx A(x) \rho (x,t) 
&=& \langle \1 | \hat{A} 
U(t,t_i)
| \rho (t_i) \rangle \nonumber \\
&=& \langle \1 | U (t_f,t)
\hat{A} 
U(t,t_i)
| \rho (t_i) \rangle. \label{eqn:exp-markov}
\end{eqnarray}
Here we used $U^\dagger (t_f,t) | \1 \rangle =  | \1 \rangle$ 
due to Eq.\ (\ref{eqn:cond-L-ad}).

Comparing with the result in the reciprocal process, 
Eq.\ (\ref{eqn:exp-markov}) is found to coincide 
with Eq.\ (\ref{eqn:exp-rp}) when the following conditions are satisfied;
1) we use the same $\hat{\cal L}_t$ satisfying Eqs.\ (\ref{eqn:l-ld}) and (\ref{eqn:cond-L-ad}) 
to express $f_{1|1}$ for both processes, and 
2) the wave functions are identified as  
\begin{eqnarray} 
| \rho (t_i) \rangle &=& K | \underline{\theta} (t_i) \rangle, \label{eqn:cond1}\\ 
| \1 \rangle &=& \frac{1}{K} | \overline{\theta} (t_f) \rangle , \label{eqn:cond2}
\end{eqnarray}
with a real constant $K$.
These conditions are realized for the transition from an initial equilibrium state. 
It is because the equilibrium state in the Markov process is given by 
$\hat{\cal L}_{t_i} | \rho_{eq}(t_i) \rangle= 0$ and this corresponds to  
$ | \underline{0} ,t_i \rangle$ with $\bar{\lambda}_0 = 0$. See Ref.\ \cite{koide17} for details. 
Then the initial conditions of the wave functions should be chosen by
\begin{eqnarray}
| \underline{\theta} (t_i) \rangle = | \underline{0} ,t_i \rangle 
&\longrightarrow& 
| \rho_{eq}(t_i) \rangle = K | \underline{0} ,t_i \rangle , \\
| \overline{\theta} (t_i) \rangle = | \overline{0} , t_i \rangle 
&\longrightarrow& 
 | \1 \rangle = \frac{1}{K} | \overline{0} , t_i \rangle . \label{eqn:cond2-2}
\end{eqnarray} 
Because of Eq.\ (\ref{eqn:cond-L-ad}) and $\langle x | \overline{0}, t_i  \rangle = const.$, 
we find $\hat{\cal L}^\dagger_{t'} | \overline{0} ,t_i \rangle =0$, leading to 
$| \overline{0} ,t \rangle = | \overline{0} ,t_i \rangle$ for any $t$.
Therefore Eq.\ (\ref{eqn:cond2}) is satisfied for this initial condition.
In short, the expectation values with $\langle \1 |$ and $| \rho_{eq} (t_i) \rangle$ 
in the Markov process is equivalent to those with $\langle \overline{0}, t_i |$ (or equivalently $\langle \overline{0}, t_f |$) and 
$| \underline{0}, t_i \rangle$ in the reciprocal process.

Similarly, the expectation values even with the joint probability densities agree in both processes 
under this initial condition.
In fact, Eq.\ (\ref{eqn:thetam}) is satisfied as 
\begin{eqnarray}
\langle \overline{0}, t_f | U(t_f,t_i) | \underline{0}, t_i \rangle 
=1.
\end{eqnarray}
Therefore we can choose 
\begin{eqnarray}
| \underline{m}, t_i \rangle &\longrightarrow& | \underline{0} , t_i \rangle, \\
| \overline{\theta}_m (t_f) \rangle &\longrightarrow& | \overline{0} , t_f \rangle,
\end{eqnarray}
in Eq.\ (\ref{eqn:gr}) and the relation for the reciprocal process characterizes even the property in the Markov process.

As an example, we choose the Fokker-Planck operator as $\hat{\cal L}_t$, 
\begin{eqnarray}
&& \langle x | \hat{\cal L}_t | x^\prime \rangle = {\cal L}_t(x) \delta (x-x^\prime), \\
&& {\cal L}_t (x)= \frac{1}{\nu\beta}\partial^2_x + \frac{1}{\nu}\partial_x V^{(1)}(x,t), 
\end{eqnarray}
where $\nu$ is the constant friction coefficient and $V^{(1)}(x,t) = \partial V(x,t)/\partial x$. The external confinement potential $V$ is time-dependent because the form is controlled by a time-dependent external parameter.
As is shown in Ref.\ \cite{koide17}, $\bar{\lambda}_n (t)\ge 0$ for integers $n \ge 0$ 
satisfying $\bar{\lambda}_0 (t) = 0$ in this case.
This generator further satisfies the TG reciprocal symmetry (\ref{eqn:l-ld}) by choosing
\begin{eqnarray}
G_t (x) = \beta H_t(x) = \beta V(x,t), \label{eqn:g}
\end{eqnarray}
where $\beta = 1/(k_B T)$ with $T$ being temperature.
Then we consider the transition from the equilibrium state defined by  
\begin{eqnarray}
\langle x | \rho_{eq} (t_i) \rangle = K \langle x | \underline{0} ,t_i \rangle = \frac{e^{-G_{t_i}(x)}}{Z(t_i)}, 
\end{eqnarray}
with the partition function $Z(t) = \int dx\ e^{-G_t (x)}$.

In the derivation of the Jarzynski relation, 
we consider a thermodynamic process realized by changing the form of the potential $V(x,t)$, and then 
the fluctuating performed work $W$ is observed \cite{jar97}. 
The expectation value for this work distribution, 
which is denoted by $\langle e^{-\beta W} \rangle$,  
is defined by using Eq.\ (\ref{eqn:gr}), leading to
\begin{eqnarray}
\langle e^{-\beta W} \rangle
&\equiv&
\langle \overline{0} ,t_f | e^{-\Delta G} | \underline{0} ,t_i \rangle_{jt} \nonumber \\
&=& 
\langle \overline{0} ,t_f | e^{-\hat{G}_{t_f}} | \overline{0} ,t_i \rangle 
\langle \underline{0} ,t_i | e^{\hat{G}_{t_i}} | \underline{0} ,t_i \rangle \nonumber \\
&=& 
e^{-\beta (F(t_f) - F(t_i))},
\end{eqnarray}
where $\hat{\cal L}^\dagger_t | \overline{0}, t_i \rangle = 0$ and the Helmholtz free energy is 
\begin{eqnarray}
F(t) = -\frac{1}{\beta} \ln Z(t).
\end{eqnarray}
This characterizes the relation between 
the change of the free energy and the distribution of the work and is known as 
the Jarzynski relation.

There is a remark for the initial condition.
In quantum mechanics, any eigenstate of a Hamiltonian can 
be used as an initial state, but it is not the case for the present calculation.
Suppose that there is a state $|\rho (t) \rangle$ normalized by one, $\int dx \langle x | \rho(t)\rangle = 1$, 
and we expand it as
${\displaystyle | \rho (t)\rangle = \sum_{n=0}^\infty c_n(t) | \underline{n},t \rangle}$,
where the coefficient is $c_n (t)= \langle \overline{n},t | \rho (t) \rangle$.
For such a state, $c_0 (t)$ is always finite because $\langle \overline{0},t | x \rangle = const$.
That is, any physical initial state must have the contribution from the component of $n=0$, differently from quantum mechanics. 
In addition, from Eq.\ (\ref{eqn:normal}), 
we can show the following property, 
$\int dx \langle x | \underline{n}, t\rangle = 0$ for $n > 0$.

\section{Symmetry in Interaction Picture}\label{sec:5}

The TG reciprocal symmetry (\ref{eqn:l-ld}) is sometimes not manifest.
Let us consider the Kramers equation, which has the generator, 
\begin{eqnarray}
{\cal L}_t (x,p) 
= 
{\cal L}^0_t (x,p) 
+
{\cal L}^1_t (x,p), 
\end{eqnarray}
where, using the particle mass $m$,  
\begin{eqnarray}
{\cal L}^0_t (x,p) 
&=& 
-\frac{p}{m} \partial_x + V^{(1)}(x,t) \partial_p, \\ 
{\cal L}^1_t (x,p) 
&=& 
 \partial_p \frac{\nu }{m} p  
+ \frac{\nu}{\beta} \partial^2_p.
\end{eqnarray}
Note that ${\cal L}^0_t$ is anti-self-adjoint because ${\cal L}^0_t = -i L^0_t$ with $L^0_t$ being the self-adjoint Liouville operator.
Differently from the Fokker-Planck equation, the matrix elements are calculated with the basis, 
$| x,p \rangle = | x \rangle \otimes | p \rangle$, where 
$| p \rangle$ satisfies the same properties as $| x\rangle$ given by Eq.\ (\ref{eqn:x}). 
Choosing
\begin{eqnarray}
G_t (x,p) = \beta {H}_t (x,p) = \beta \left( \frac{p^2}{2m} + V(x,t) \right),
\end{eqnarray}
the transformation law of the generator is given by 
\begin{eqnarray}
e^{\hat{G}_t} \hat{\cal L}_t e^{-\hat{G}_t}
= -(\hat{\cal L}^0_t)^\dagger + (\hat{\cal L}^1_t)^\dagger, \label{eqn:hidden}
\end{eqnarray}
and Eq.\ (\ref{eqn:l-ld}) is not satisfied.

The TG reciprocal symmetry of this system is hidden.
To see it, we introduce the operator and the state in the ``interaction" picture by
\begin{eqnarray}
\hat{A}_{I}(t) &\equiv& U^\dagger_0 (t) \hat{A}_t U_0 (t), \\
| \underline{\theta}_I (t) \rangle &\equiv& U^\dagger_0 (t) | \underline{\theta} (t) \rangle,
\end{eqnarray}
with 
\begin{eqnarray}
U_0 (t) &=& T e^{\int^t_{-\infty} ds \hat{\cal L}^0_s}, \\ 
U^\dagger_0(t) &=& U^{-1}_0 (t).
\end{eqnarray}
The time evolutions in this picture are  
\begin{eqnarray}
\partial_t | \underline{\theta}_I (t) \rangle  
&=& \hat{\cal L}^1_I(t)  | \underline{\theta}_I (t) \rangle, \\ 
\partial_t \langle \overline{\theta}_I (t) |  
&=& \langle \overline{\theta}_I (t) | (-\hat{\cal L}^1_I(t)).
\end{eqnarray}
We can see that this generator in the interaction picture satisfies the TG reciprocal symmetry, 
\begin{eqnarray}
e^{\hat{G}_I(t)} \hat{\cal L}^1_I(t) e^{-\hat{G}_I(t)} = (\hat{\cal L}^1_I(t))^\dagger. \label{eqn:l-ld2}
\end{eqnarray}

The form of $\hat{\cal L}^1_t$ is the same as that of the Fokker-Planck operator and 
we can use the same properties for the eigenstates,
\begin{eqnarray}
\hat{\cal L}^1_I(t) | \underline{n}_I, t \rangle &=& -\bar{\lambda}^{1}_n (t) | \underline{n}_I, t \rangle, \\
(\hat{\cal L}^1_I(t))^\dagger | \overline{n}_I, t \rangle &=& -\bar{\lambda}^{1}_n (t) | \overline{n}_I, t \rangle,
\end{eqnarray}
with $\bar{\lambda}^1_0 (t) = 0$.
The eigenstate of $\hat{\cal L}^1_I(t)$ satisfies the similar relation to 
Eq.\ (\ref{eqn:wf-relation1}), 
\begin{eqnarray}
e^{\hat{G}_I(t)}| \underline{n}_I, t \rangle = {\cal N}^{I}_n (t) | \overline{n}_I, t \rangle,
\end{eqnarray}
with the prefactor ${\cal N}^{I}_n(t)$.
Then Eq.\ (\ref{eqn:gr}) is reexpressed in the interaction picture as 
\begin{eqnarray}
\lefteqn{\langle \overline{\theta}_m  (t_f) | e^{-\Delta G} | \underline{m} (t_i) \rangle_{jt}} && \nonumber \\ 
&=& 
\langle (\overline{\theta}_{I})_m (t_f) | e^{-\hat{G}_I({t_f})} 
T e^{\int^{t_f}_{t_i} ds (\hat{\cal L}^1_I(s))^\dagger } 
| \overline{m}_I ,t_i \rangle \nonumber \\
&& \times \langle \underline{m}_I ,t_i | e^{\hat{G}_I(t_i)} | \underline{m}_I ,t_i \rangle. 
\label{eqn:gr2}
\end{eqnarray}
Here we used 
\begin{eqnarray}
\hspace*{-0.5cm}
U^\dagger_0 (t_{M+1}) 
U(t_{M+1},t_M)U_0 (t_M)
&=&  
T e^{\int^{t_{M+1}}_{t_M}ds \hat{\cal L}^1_I(s)}. 
\end{eqnarray}

To express the transition from an initial equilibrium state in the Markov process, 
we should choose the initial conditions for the wave functions by 
\begin{eqnarray}
| \rho_{eq} (t_i) \rangle &=& K | \underline{0} ,t_i \rangle, \\
| \1 \rangle &=& \frac{1}{K} | \overline{0} ,t_i \rangle,
\end{eqnarray}
where $| \underline{0} ,t_i \rangle$ and $| \overline{0} ,t_i \rangle$ are 
zero eigenstates of $\hat{\cal L}_{t_i}$ (also $\hat{\cal L}^1_{t_i}$) and $\hat{\cal L}^\dagger_{t_i}$ 
(also $(\hat{\cal L}^1_{t_i})^\dagger$), respectively.
Therefore the corresponding states in the interacting picture are 
\begin{eqnarray}
| (\rho_{eq})_I (t_i) \rangle &=& K | \underline{0}_{I} ,t_i \rangle, \\
| \1_{I} \rangle &=& \frac{1}{K} | \overline{0}_{I} ,t_i \rangle.
\end{eqnarray}

Because $(\hat{\cal L}^1_I(t))^\dagger | \overline{0}_I ,t_i \rangle = 0$, Eq.\ (\ref{eqn:gr2}) leads to 
the Jarzynski relation for the Kramers equation,
\begin{eqnarray}
\langle e^{-\beta W}  \rangle 
&=&
\langle \overline{0}_I ,t_f | e^{-\hat{G}_I({t_f})} | \overline{0}_I ,t_i \rangle  
\langle \underline{0}_I ,t_i | e^{\hat{G}_I({t_i})} | \underline{0}_I ,t_i \rangle \nonumber \\
&=&
e^{-\beta (F(t_f) - F(t_I) )}.
\end{eqnarray}
The result is independent of the value of $\nu$. If we consider the vanishing limit of $\nu$, 
it represents the result by the evolution with the Liouville operator itself.

This discussion is easily applicable to the relativistic Kramers equation \cite{koide-rbm,hanggi09}, 
\begin{eqnarray}
{\cal L}^0_t (x,p) 
&=& 
 - \frac{p}{p^0} \partial_x+ V^{(1)}(x,t) \partial_p , \\
{\cal L}^1_t (x,p) 
&=& 
 \partial_p \frac{\nu }{p^0} p  
+ \frac{\nu}{\beta} \partial^2_p,
\end{eqnarray}
where $p^0 = c\sqrt{p^2 + m^2 c^2}$.
This generator has the same TG reciprocal symmetry as that of the Kramers equation by defining 
$H_t(x,p)= c\sqrt{p^2 + m^2 c^2} + V(x,t)$. We again confirm the Jarzynski relation
\footnote{
Note that, because of the existence of the heat bath and the introduction of the external potential, 
the relativistic Kramers equation is not Lorentz-covariant and written as the equation for the rest frame of the heat bath. 
}.

In stochastic systems, the form of the generator can depend on the definition of the stochastic integral. 
For example, in Ref.\ \cite{koide-rbm}, three different equations are obtained using the Ito, 
Stratonovich-Fisk and H\"{a}nggi-Klimontovich definitions. 
The Jarzynski relation is satisfied independently of the choice of the definitions.

\section{Concluding remarks}\label{sec:6}

In this paper, we showed that the Jarzynski relation is the realization of the TG reciprocal symmetry 
(\ref{eqn:l-ld}), which is related to the invariance of the Lagrangian and sometimes hidden.
We further introduced the reciprocal process, showing 
that the descriptions by the Markov process from an initial equilibrium state 
are indistinguishable from those by the reciprocal process. 
Then the mathematical relation satisfied for the reciprocal process, Eq.\ (\ref{eqn:gr}) (or (\ref{eqn:gr2})),
describes even the behavior of the Markov process. 
Finally we showed that the Jarzynski relation is reproduced from the derived mathematical relation 
when it is applied to the Fokker-Planck, Kramers and relativistic Kramers equations.
In principle, the reciprocal process can be realized experimentally by the method explained in Ref.\ \cite{eqm1}. 
Then Eq.\ (\ref{eqn:gr}) (or (\ref{eqn:gr2})) will be confirmed in such an experiment.

So far we have considered a constant temperature, but the discussions are applicable 
even when the temperature has a time dependence. Then the Jarzynski relation is modified as 
$\langle e^{-\beta W} \rangle= e^{- (\beta(t_f) F(t_f) - \beta(t_i) F(t_i))}$.

The property for the time generator analogous to our symmetry is considered by Eq.\ (3.15) in Ref.\ \cite{kurchan07}.
However, because the time dependences of $\eta_t$ and $\hat{\cal L}_t$ are omitted in the definitions there,   
it is not clear how the result in Ref.\ \cite{kurchan07} is reproduced in the present framework. 
For example, we can choose 
$\hat{\eta}_t = {\cal T} e^{\hat{G}_t}$
with the time reversal operator ${\cal T}$ to reproduce Eq.\ (3.15) of Ref.\ \cite{kurchan07}. 
To obtain the condition corresponding to Eq.\ (\ref{eqn:wf-relation1}), 
however, Eq.\ (\ref{eqn:l-ld}) will be modified as 
$\hat{\eta}_t \hat{\cal L}_t \hat{\eta}^{-1}_t = \hat{\cal L^\dagger}_{t'}$
with $t' = t_i + t_f -t$.

It is known that the Jarzynski relation can be obtained in the Markov process and thus 
it is not necessary to introduce the reciprocal process for the derivation.
In fact, if we choose the boundary joint probability density by the following form, 
the reciprocal process coincides with the Markov process, 
\begin{eqnarray}
c_{0}(x_f,t_f;x_i,t_i) = \langle x_f | U(t_f,t_i) | x_i \rangle \langle x_i | \rho (t_i) \rangle.
\end{eqnarray}
This choice is practically equivalent to choose $\langle \overline{\theta}(t) | \propto \langle \1 |$ in Eq.\ (\ref{eqn:entangle}).
However, 
the reciprocal process is essential to consider the Jarzynski relation from the perspective of 
the TG reciprocal symmetry. 
Then, $| \1 \rangle$ is transformed to the final equilibrium state by 
operating $e^{-\hat{G}_{t_f}}$, independently of the behavior of $| \rho(t_f) \rangle$.
This is the mathematical reason why 
the final equilibrium free energy appears in the Jarzynski relation even if the final state is not limited to be an equilibrium state. See also the discussion in Ref.\ \cite{chet11}.

Note that Eq.\ (\ref{eqn:gr}) represents the expectation value with the joint probability density fixing the initial and final states. 
However, it should be noted that, when the generator satisfies Eq.\ (\ref{eqn:cond-L-ad}), what is practically fixed is only the initial state even in the reciprocal process. 
In fact, there are two time evolutions of the states, $|\underline{\theta}(t)\rangle$ and 
$|\overline{\theta}(t) \rangle$, but only $|\overline{\theta}(t_f) \rangle$ is fixed in Eq.\ (\ref{eqn:gr}). 
Moreoever,  as we showed below Eq.\ (\ref{eqn:cond2-2}), the evolution of 
$|\overline{\theta}(t) \rangle$ is trivial, $|\overline{\theta}(t_i) \rangle = |\overline{\theta}(t_f) \rangle$.
Therefore the reciprocal process practically fixes only the 
initial state $|\underline{\theta}(t_i)\rangle$ as is the standard Markov process.

Because the framework of the reciprocal process is more general than the Markov process, 
it is possible to consider the classical process where we have to take into account 
the non-trivial time dependence of $\langle \overline{\theta}(t) |$, 
differently from the discussions so far. 
Moreover, 
the boundary joint probability density in this case has 
a non-trivial correlation which remains us of quantum mechanics, and 
it is interesting to imagine  
that a quantum-mechanical behavior, such as entanglement, is observed under a special setup 
in the classical statistical physics. 
For example, the motion of a droplet on the surface of a liquid shows the behaviors close to quantum mechanics 
\cite{couder06,couder09,brandy14,bush15} and this may be understood as the reciprocal process.
See also discussions in Refs.\ \cite{eqm1,eqm2,yasue,vigier}.

The reciprocal process has a similar mathematical structure to quantum mechanics, 
and the present derivation of the Jarzynski relation will be useful to deepen our understanding for 
the connection between
the Jarzynski relations in classical and quantum systems.
Then it is expected that Eq.\ (\ref{eqn:gr}) leads to the quantum Jarzynski relation when it is generalized to the complex Hilbert space eliminating the final state by using the relation  
$\langle \overline{\theta}(t_f)| = \langle \overline{\theta}(t_i)|U^{-1}(t_f,t_i)$. 
If the unified description of the Jarzynski relations is possible, it is interesting to ask whether 
the TG reciprocal symmetry in this paper is still preserved even in quantum systems.
In Ref.\ \cite{deffner15}, the quantum Jarzynski relation is studied in the association with 
the ${\cal PT}$-symmetry of the Hamiltonian. 
The TG reciprocal symmetry (\ref{eqn:l-ld}) can be regarded as the so-called pseudo-Hermiticity \cite{mostafa02} and the ${\cal PT}$-symmetry is known to be 
a special case of the pseudo Hermiticity \cite{bender07}. 
Our generators are however not necessarily ${\cal PT}$-symmetric and the TG reciprocal symmetry is sometimes hidden. 
This difference should be understood.

\vspace{1cm}

The author acknowledges to the referees of Ref.\ \cite{koide17}, because this work started out as an answer to 
their comments. Thanks also to C.\ E.\ Aguiar, V.\ A.\ S.\ V.\ Bittencourt, M.\ Blasone, V.\ Brasil, H.-T.\ Elze, 
T.\ Kodama and K.\ Tsushima for useful comments and references.
This work is financially supported by Conselho Nacional de
Desenvolvimento Cient\'{i}fico e Tecnol\'{o}gico (CNPq), project 307516/2015-6, 
and a part of the project INCT-FNA Proc.\ No.\ 464898/2014-5.

\appendix

\section{TG reciprocal symmetry and invariance of Lagrangian} \label{app:t-reciprocal}

We consider the system of the damped and amplified harmonic oscillators, 
\begin{eqnarray}
&& \left( \frac{d^2}{dt^2} + \frac{\gamma}{m}\frac{d}{dt} + \omega^2 \right) x = \hat{\cal L} x = 0, \label{eqn:app1}\\
&& \left( \frac{d^2}{dt^2} - \frac{\gamma}{m}\frac{d}{dt} + \omega^2 \right) y = \hat{\cal L}^{c} y = 0, \label{eqn:app2}
\end{eqnarray}
where $\omega$ is the angular frequency and $\gamma$ denotes the dissipative coefficient.
Although the definition of the self-adjoint operation is not clear in this particle system, 
we interpret that $\hat{\cal L}^c$ is the conjugate operator of $\hat{\cal L}$.
Then we find a kind of TG reciprocal symmetry, 
\begin{eqnarray}
e^{\hat{G}_t} \hat{\cal L} e^{-\hat{G}_t} = \hat{\cal L}^c, \label{eqn:app-tsym}
\end{eqnarray}
by choosing, 
\begin{eqnarray}
\hat{G}_t = \frac{\gamma}{m}t.
\end{eqnarray}

Following Bateman \cite{bateman,blasone01,blasone04}, we introduce the Lagrangian of this system by
\begin{eqnarray}
L(x,\dot{x};y,\dot{y}) = m \dot{x} \dot{y} + \frac{\gamma}{2} (x\dot{y} - \dot{x}y) - m \omega^2 xy. \label{eqn:bate}
\end{eqnarray}
On the other hand, applying Eq.\ (\ref{eqn:app-tsym}) to Eqs.\ (\ref{eqn:app1}) and (\ref{eqn:app2}), 
we find that the variables $x_t$ and $y_t$ are exchanged by the following law, 
\begin{eqnarray}
x_t &\longrightarrow x^\prime_t =  K e^{-\hat{G}_t} y_t, \\
y_t &\longrightarrow y^\prime_t =  K^{-1} e^{\hat{G}_t} x_t.
\end{eqnarray}
Here we used that Eqs.\ (\ref{eqn:app1}) and (\ref{eqn:app2}) 
are linear and there is an ambiguity to multiply the solutions by the real constant prefactor $K$.
Then the Lagrangian is invariant for this transformation of the variables, 
\begin{eqnarray}
L(x,\dot{x};y,\dot{y}) = L(x^\prime,\dot{x}^\prime;y^\prime,\dot{y}^\prime).
\end{eqnarray}

Now we apply this argument to our problem. 
The Lagrangian density to derive the eigenvalue equations for the wave functions 
is given by 
\begin{eqnarray}
\lefteqn{L (\underline{\theta},\partial \underline{\theta};\overline{\theta},\partial \overline{\theta})} && \nonumber \\
&=& 
-\bar{\lambda} \overline{\theta}\underline{\theta}
+
\frac{1}{\nu \beta} (\partial_x \overline{\theta})  (\partial_x \underline{\theta}) 
+
\frac{1}{2\nu} \left\{
\underline{\theta} V^{(1)} \partial_x \overline{\theta}
-
\overline{\theta} \partial_x (V^{(1)} \underline{\theta})   
\right\}, \nonumber \\
\end{eqnarray}
where $\bar{\lambda}$ denotes the eigenvalue.
One can easily confirm that this Lagrangian density leads to Eqs.\ (\ref{eqn:eigen1}) and (\ref{eqn:eigen2}) 
with the Fokker-Planck operator. 

From the TG reciprocal symmetry, Eq.\ (\ref{eqn:wf-relation1}) is obtained. Thus 
we consider the following transformation of the wave functions, 
\begin{eqnarray}
\underline{\theta}(x,t)  &\longrightarrow \underline{\theta}^\prime (x,t)  = {\cal N}_\theta (t) e^{-{G}_t(x)}\overline{\theta}(x,t), \\
 \overline{\theta}(x,t)  &\longrightarrow \overline{\theta}^\prime (x,t)  = {\cal N}^{-1}_\theta (t) e^{{G}_t(x)}
\underline{\theta}(x,t).
\end{eqnarray}
Applying these to the Lagrangian density, we find the invariance of the Lagrangian density, 
\begin{eqnarray}
L (\underline{\theta},\partial \underline{\theta};\overline{\theta},\partial \overline{\theta})
 = L (\underline{\theta}^\prime,\partial \underline{\theta}^\prime;\overline{\theta}^\prime,\partial \overline{\theta}^\prime).
\end{eqnarray}
Therefore the TG reciprocal symmetry leads to the invariance of the Lagrangian density.

Differently from the case of the harmonic oscillator (\ref{eqn:bate}), 
however, the TG reciprocal symmetry discussed in the paper 
is the property of the time generator and not that of the differential equations.
In fact, we can consider the following Lagrangian density which reproduces the differential equations (\ref{eqn:de1}) and (\ref{eqn:de2}),
\begin{eqnarray}
&&\hspace*{-1cm} L 
(\underline{\theta},\partial \underline{\theta};\overline{\theta},\partial \overline{\theta})
= 
\frac{1}{2}( \overline{\theta} \partial_t \underline{\theta} -  \underline{\theta} \partial_t \overline{\theta})  \nonumber \\
&&\hspace*{-1cm} + 
\frac{1}{\nu \beta} (\partial_x \overline{\theta})  (\partial_x \underline{\theta}) 
+
\frac{1}{2\nu} \left\{
\underline{\theta} V^{(1)} \partial_x \overline{\theta}
-
\overline{\theta} \partial_x (V^{(1)} \underline{\theta})   
\right\}, 
\end{eqnarray}
but, it does not have the TG reciprocal symmetry because of the 
time-dependence in $\hat{G}_t$.

\end{document}